\documentclass{pasj00}
\def\e{\mathrm{e}}

\begin{document}
\SetRunningHead{Y. Suwa}{From supernovae to neutron stars}
%\Received{}%{yyyy/mm/dd}
%\Accepted{}%{yyyy/mm/dd}
%\Published{}%{yyyy/mm/dd}

\title{From supernovae to neutron stars}

%%% begin:list of authors
% Do NOT capitalize all letters in "textsc".
\author{Yudai \textsc{Suwa} %
%  \thanks{Example: Present Address is xxxxxxxxxx}
  }
\affil{Yukawa Institute for Theoretical Physics, Kyoto
  University, Oiwake-cho, Kitashirakawa, Sakyo-ku, Kyoto, 606-8502,
  Japan}
\email{suwa@yukawa.kyoto-u.ac.jp}

%%% end:list of authors

%%% Please use the following style in case that sorting by 
%%% affiliation is impossible. 
%
% \author{%
%   D-Firstname \textsc{D-Familyname}\altaffilmark{1}
%   E-Firstname \textsc{E-Familyname}\altaffilmark{1,2}
%   and
%   F-Firstname \textsc{F-Familyname}\altaffilmark{2}}
% \altaffiltext{1}{Address of Institute}
% \email{ddddd@xxx.xxx.xx.xx}
% \email{eeeee@xxx.xxx.xx.xx}
% \altaffiltext{2}{Address of Institute}

%% `\KeyWords{}' always has to be placed before `\maketitle'.
 %Do NOT move this preamble from here!

\maketitle

\begin{abstract}
The gravitational collapse, bounce, the explosion of an iron core of
an 11.2 $M_{\odot}$ star is simulated by two-dimensional
neutrino-radiation hydrodynamic code. The explosion is driven by the
neutrino heating aided by multi-dimensional hydrodynamic effects such
as the convection. Following the explosion phase, we continue the
simulation focusing on the thermal evolution of the protoneutron star
up to $\sim$70 s when the crust of the neutron star is formed using
one-dimensional simulation.  We find that the crust forms at
high-density region ($\rho\sim10^{14}$ g cm$^{-3}$) and it would
proceed from inside to outside. This is the first self-consistent
simulation that successfully follows from the collapse phase to the
protoneutron star cooling phase based on the multi-dimensional
hydrodynamic simulation.
\end{abstract}

%%%%%%%%%%%%%%%%%%%%%%%%%%%%%%%%%%%
%%%%%%%%%%%%%%%%%%%%%%%%%%%%%%%%%%%
\section{Introduction}
\label{sec:intro}
%%%%%%%%%%%%%%%%%%%%%%%%%%%%%%%%%%%
%%%%%%%%%%%%%%%%%%%%%%%%%%%%%%%%%%%

Core-collapse supernovae are transitions of massive stellar cores into
neutron stars (hereafter NSs).  These phenomena are significantly
dynamical events, since the central density increases by $\sim 10^{6}$
(from $\sim 10^{9}$ g cm$^{-3}$ to $\sim 10^{15}$ g cm$^{-3}$) and the
radius decreases by $\sim$ 100 (from $\sim$1000 km to $\sim$ 10 km)
within only $\sim 1$ second.

A NS can be divided into two components; the (superfluid) core and the
crust.  In the crust, Coulomb forces dominate the thermal fluctuation
and nuclei crystallize into a periodic (body centered cubic) lattice
structure that minimizes the Coulomb energy.
The presence of the crust implies typical aspects of NSs. For example,
the cooling curve of old pulsars is characterized by the heat
conductivity in the crust (see \cite{yako04} for review and references
therein), and the giant flares of soft-gamma ray repeaters are
conjectured to originate from the sudden release of the magnetic field
energy, which is stored below the crust and breaks the crust when the
magnetic stress overwhelms the crustal stress \citep{thom01}.  Also
the relationship with the pulsar glitches \citep{rude98} and the
gravitational waves from mountains on the crust \citep{horo09} are
discussed.  The crust, however, does not exist just after the
supernova explosion sets in. At first, the neutron star is very hot
(the temperature $T\sim 10^{11}$ K) so that it entirely behaves as
fluid (so-called ``protoneutron star''; PNS). As the PNS cools down by
neutrino emission and at some point the thermal energy becomes
comparable to the Coulomb energy, nuclei form a lattice structure,
which corresponds to the crust formation.

Recently, by grace of the growing computer resources and development
of the numerical scheme, we have several numerical studies that
performed multi-dimensional (multi-D) hydrodynamic simulations with
neutrino radiative transfer and succeeded to make the shock expansion
up to outside the iron core (e.g.,
\cite{bura06,burr06,mare09,suwa10,taki12,muel12b,brue13}). The shock
expansion was obtained by the multi-D effects that amplify the
neutrino heating rate such as the convection and standing accretion
shock instability (SASI), even though state-of-the-art one-dimensional
(1D) simulations failed to produce the explosion
\citep{ramp00,lieb01,thom03,sumi05}. By these facts, we are now
standing at the position to be able to carry on the numerical
simulation of a NS crust formation, which can be called as the true NS
formation, by using self-consistent exploding models.

In this letter, we report the simulation result from the onset of an
iron core collapse to the formation of neutron star crust.  Though the
explosion mechanism of supernovae is still under the thick veil, in
this letter we rely on the standard scenario, i.e., the delayed
explosion scenario, in which the neutrino heating is crucial and the
neutrino transfer equation should be solved in order to estimate the
neutrino-heating rate.  In previous studies, \citet{burr86}
investigated long-term evolution of Kelvin-Helmholtz cooling phase of
the PNS starting from a handmade initial condition, which imitates the
hydrodynamical profiles after the bounce (see also \cite{pons99} for a
recent study). More recently, \citet{fisc10} and \citet{fisc12}
performed simulations with fully-general-relativistic
radiation-hydrodynamic code up to $\sim$10 seconds. Since their code
is 1D and no self-consistent explosion of an iron core can be
reproduced, they artificially amplified the charged current reaction
rate in order to produce explosion (one exemption is an $8.8M_{\odot}$
progenitor, which, however, has an O-Ne-Mg core, not an iron core).
Therefore, this letter represents the first study using the
self-consistent neutrino-radiation hydrodynamic simulation in multi-D,
which successfully follows the very long-term evolution of the massive
stellar core; the core collapse of an iron core, the bounce and the
shock formation at the nuclear density, the shock expansion, the
neutrino-driven wind phase, and finally PNS cooling phase. In
addition, this letter investigates when the crust forms inside the
PNS, which is studied for the first time using core-collapse supernova
simulation.

In this letter, \S2 is devoted to the explanation of numerical
methods. In \S3, we give our results of simulation, focusing on the
thermal evolution of PNS. We summarize our results and discuss their
implications in \S4.

%%%%%%%%%%%%%%%%%%%%%%%%%%%%%%%%%%%
%%%%%%%%%%%%%%%%%%%%%%%%%%%%%%%%%%%
\section{Numerical Method}
\label{sec:method}
%%%%%%%%%%%%%%%%%%%%%%%%%%%%%%%%%%%
%%%%%%%%%%%%%%%%%%%%%%%%%%%%%%%%%%%

Numerical methods are basically the same as ones in our previous
papers \citep{suwa10,suwa11b,suwa13b}. With the ZEUS-2D code
\citep{ston92} as a base for the solver of hydrodynamics, we employ an
equation of state (EOS) of \citet{shen98}, which is able to reproduce
a $1.97 M_\odot$ NS of \citet{demo10}, and solve the spectral neutrino
radiative transfer equation for $\nu_\e$ and $\bar\nu_e$ using
isotropic diffusion source approximation (IDSA), which is well
calibrated to reproduce the results of full Boltzmann solver
\citep{lieb09}.  The weak interaction rates for neutrinos are
calculated by using the formulation of \citet{brue85}.  The
simulations are performed on a grid of 300 logarithmically spaced
radial zones up to 5000 km with the least grid width being 1 km at the
center and 128 equidistant angular zones covering $0<\theta<\pi$ for
two-dimensional (2D) simulation. As for the continuous 1D simulation,
the same radial zoning is used and ``one'' angular grid is employed.
In order to resolve the steep density gradient at the PNS surface and
remove the outermost region with the density $\rho<10^{5}$ g
cm$^{-3}$, which is the lowest value of EOS table, we perform rezoning
several times for later 1D run. At last, the radius of the minimum
grid (and the least grid width) is 300 m and the maximum grid edge
locates at 100 km from the center. The total mass of PNS is conserved
over these rezoning within $\sim$0.1\% error.  For neutrino transport,
we use 20 logarithmically spaced energy bins reaching from 3 to 300
MeV.  As for the initial condition, we employ an 11.2 $M_{\odot}$
model from \citet{woos02}, with which several papers reported the
successful explosion \citep{bura06, mare09, taki12, suwa13b}.

%%%%%%%%%%%%%%%%%%%%%%%%%%%%%%%%%%%
%%%%%%%%%%%%%%%%%%%%%%%%%%%%%%%%%%%
\section{Results}
\label{sec:result}
%%%%%%%%%%%%%%%%%%%%%%%%%%%%%%%%%%%
%%%%%%%%%%%%%%%%%%%%%%%%%%%%%%%%%%%

The basic picture of core-collapse supernova is following
\citep{beth90,jank12,kota12}; i) the collapse of an iron core driven
by the energy loss due to the photodissociation of iron and the
electron capture, ii) the neutron star formation and the bounce shock
production, iii) the shock stall due to the neutrino cooling, iv) the
shock revival by the neutrino heating, v) the neutrino-driven wind
phase, and vi) the neutron star cooling phase. The previous
hydrodynamic studies have focused only on early phases, mostly the
shock revival process, which is shorter than $\sim 1$ s after the
bounce (by phase iv; but see \cite{fisc10} for the phase v). The PNS
cooling has much longer timescale $\sim O(10)$ s so that fully
consistent simulation has not been done so far.

Here, we perform two-dimensional simulation of the collapse, bounce,
and the onset of the explosion by the neutrino heating up to 690 ms
after the bounce.  The core bounce occurs 150 ms after the simulation
starts with the central density of $\approx 3.1\times 10^{14}$ g
cm$^{-3}$.  Soon after the bounce, the shock propagates outside the
neutrinosphere, the convection sets in and the convective plumes hit
the shock.  In this simulation, the shock does not stall and the
successful explosion occurs (see \cite{suwa13b} for more details).
After that all hydrodynamic quantities are averaged over the
angle\footnote{This treatment does not produce any strange phenomenon
  to the PNS because the PNS is almost spherically symmetric for the
  case without the rotation.} and the spherically symmetric simulation
is followed up to $\sim$ 70 s when the crust formation condition is
satisfied.  Note that the whole simulation is performed using the
``same'' code so that there is no discontinuity between these 2D and
1D simulations. If we use the different codes to connect the different
times and physical scales, there should occur some breaks of physical
quantities (e.g., total mass, total momentum, and total
energy). Therefore, the consistent simulation with the same code has
advantage for the disappearance of these breaks.

\begin{figure}[tbp]
\centering
\includegraphics[width=0.45\textwidth]{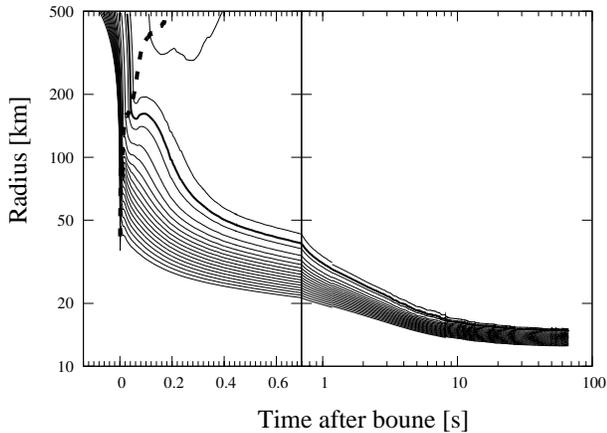}
\caption{Trajectories of selected mass coordinates from 1.01
  $M_{\odot}$ to 1.33 $M_{\odot}$ by a step of 0.02 $M_{\odot}$. The
  thick solid line indicates the position of $1.3 M_{\odot}$, which
  indicates the mass of the PNS, and the thick dotted line represents
  the shock radius at the northern pole. The left panel is the result
  of 2D simulation and the right panel is that of continuous 1D
  simulation with the connection done on $\sim 690$ ms after the
  bounce. The shrinkage of the PNS can be seen.  There are several
  discontinuities, for example $\sim$1.2 s postbounce, which are due
  to the rezoning to make resolution finer and remove the outermost
  region where the density becomes too small to use the tabular
  equation of state. This discontinuities do not make any serious
  problems in this simulation.  }
\label{fig:mass_shell}
\end{figure}

In figure \ref{fig:mass_shell}, the time evolution of selected mass
coordinates is presented. The mass within $\sim 1.3 M_{\odot}$
contracts to a PNS and the outer part expands as an ejecta of the
supernova. The shock (thick dotted line) propagates rapidly to outside
the iron core driven by the neutrino heating aided by the convective
fluid motion. The estimated diagnostic energy \citep{suwa10}
(so-called the {\it explosion energy}) determined by summing up the
gravitationally unbound fluid elements in this model is $\sim 10^{50}$
erg so that the {\it realistic} explosion simulation is still not
achieved. This is, however, one of {\it successful} explosion
simulations. The term ``successful'' means that the simulation
successfully reproduces the structure containing a remaining PNS and
escaping ejecta. Previous exploding models obtained by \citet{mare09}
and \citet{suwa10} certainly acquired the expanding shock wave up to
outside the iron core, but most of postshock materials were infalling
so that the mass accretion onto PNS did not settle and the mass of the
PNS continued to increase. Thus these simulations were {\it not fully
  successful} explosions. On the other hand, \citet{suwa13b} and this
work successfully reproduce the envelope ejection so that we can
determine the ``mass cut'', which gives the final mass of the compact
object (i.e., NS or black hole). This is because the progenitor used
in this study has a steep density gradient between iron core and
silicon layer so that ram pressure of infalling material rapidly
decreases when the shock passes the iron core surface. This is similar
situation to the explosion simulation of an O-Ne-Mg core of an 8.8
$M_{\odot}$ progenitor reported by \citet{kita06}, in which the
neutrino driven explosion was obtained in ``1D'' simulation owing to
very steep density gradient of this specific progenitor. Note that the
progenitor in this study does not explode in spherical symmetry even
though it has a steep density gradient. However, by the help of
convection, this progenitor explodes in 2D simulation and the shock
earns enough energy to blow away the outer layers.

\begin{figure*}[tbp]
\centering
\includegraphics[width=0.45\textwidth]{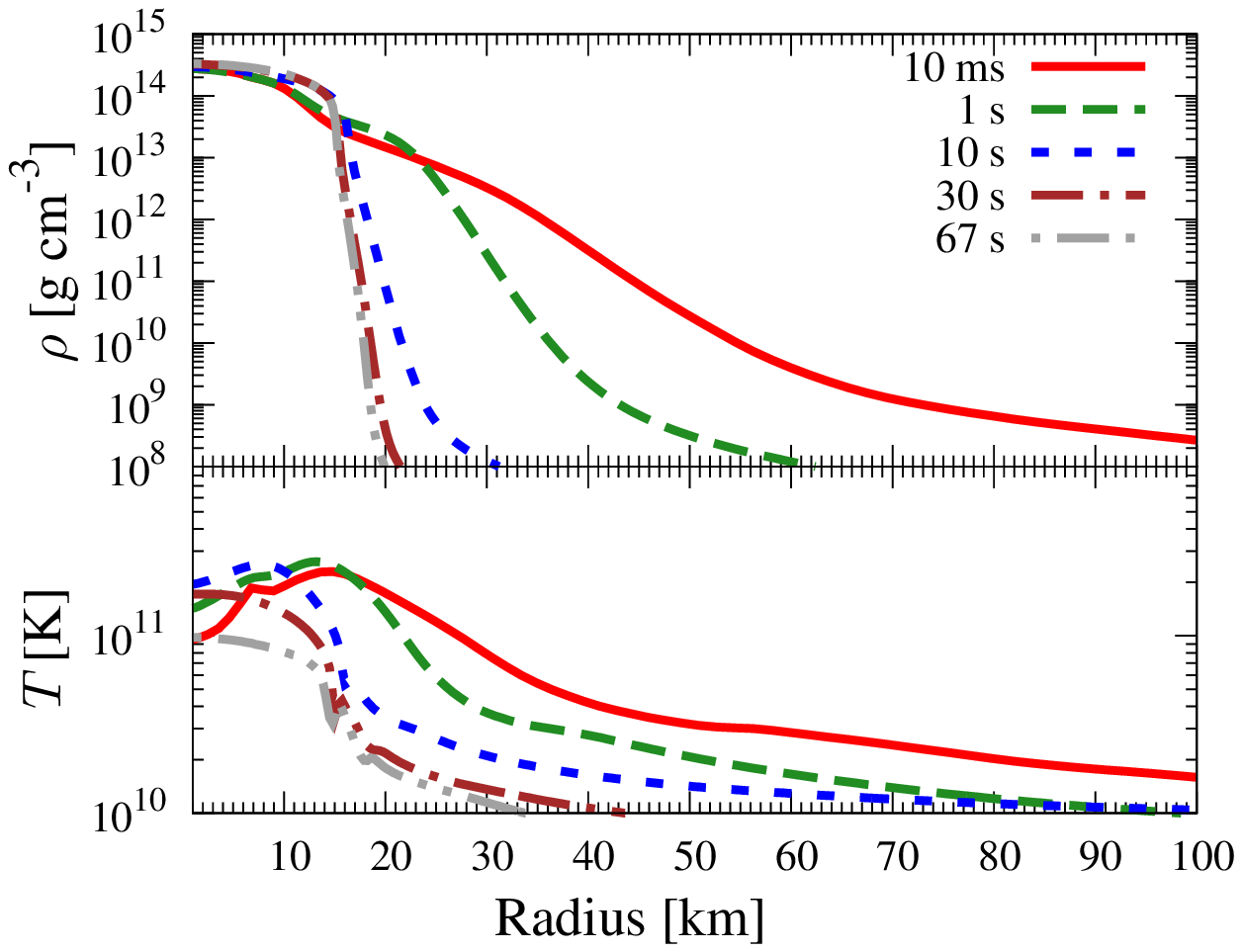}
\includegraphics[width=0.45\textwidth]{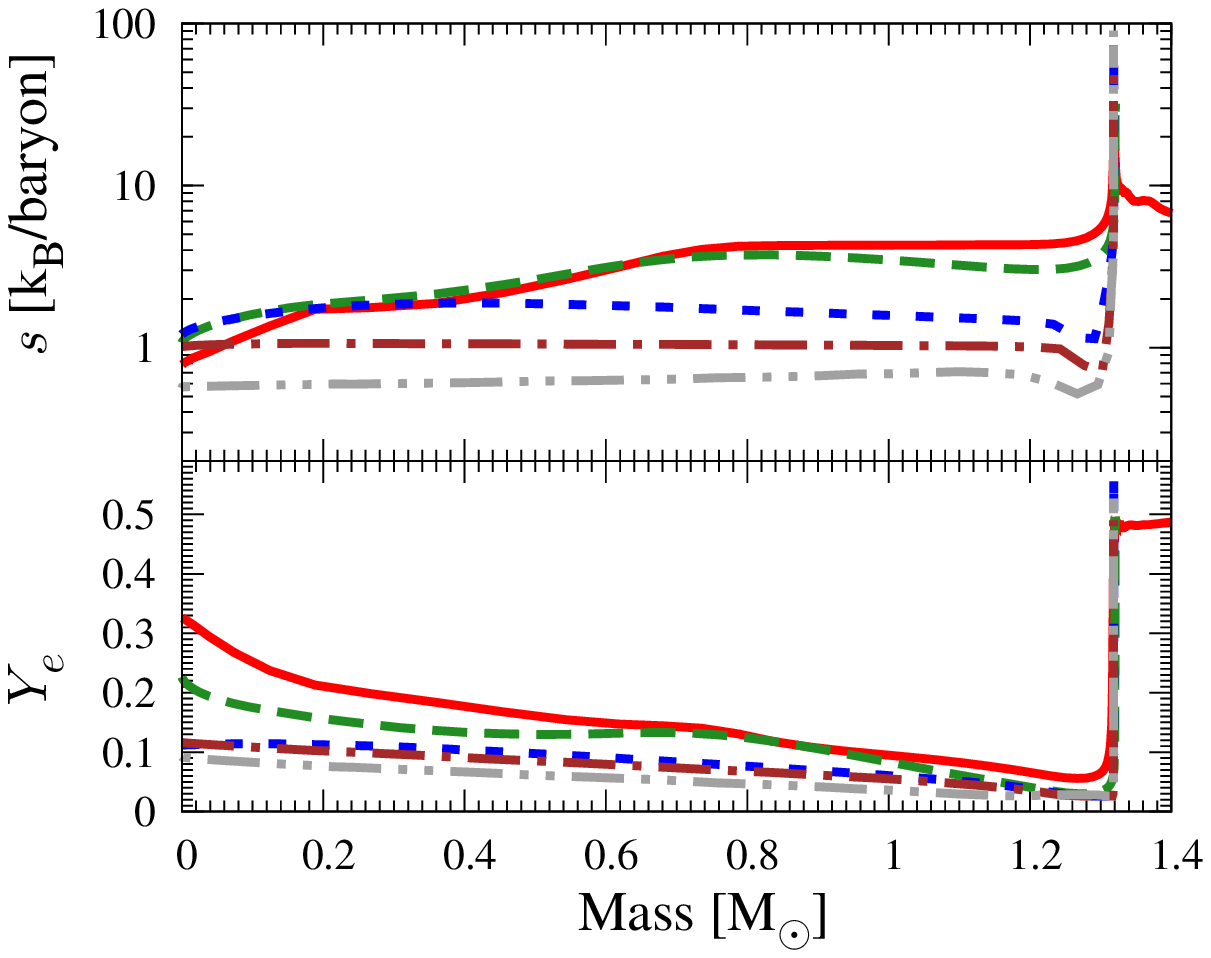}
\caption{Time evolution of the density (left top), the temperature
  (left bottom), the entropy (right top), and the electron fraction
  (right bottom). The density and the temperature are given as
  functions of the radius and the entropy and the electron fraction
  are functions of the mass coordinate. The corresponding times
  measured from the bounce are 10 ms (red solid line), 1 s (green
  dashed line), 10 s (blue dotted line), 30 s (brown dod-dashed line),
  and 67 s (grey dot-dot-dashed line), respectively.  }
\label{fig:hydro}
\end{figure*}

In figure \ref{fig:hydro}, we show hydrodynamic quantities ($\rho$,
$T$, entropy $s$, and electron fraction $Y_{e}$) at several selected
times, i.e., 10 ms, 1 s, 10 s, 30 s, and 67 s after the bounce. One
can find by the density plot that the PNS shrinks due to neutrino
cooling. Note that for the postbounce time $t_\mathrm{pb}\lesssim 10$
s the central temperature increases because of the equidistribution of
the thermal energy that can be found in the entropy plot, in which one
finds that entropy at the center increases due to entropy flow from
the outer part. For $t_\mathrm{pb}\gtrsim 10$ s, the PNS evolves
almost isentropically and both the entropy and the temperature
decrease due to the neutrino cooling. This can be called as the PNS
cooling phase.  One can find from $Y_e$ plot that neutrinos take out
the lepton number as well.

\begin{figure}[tbp]
\centering
\includegraphics[width=0.45\textwidth]{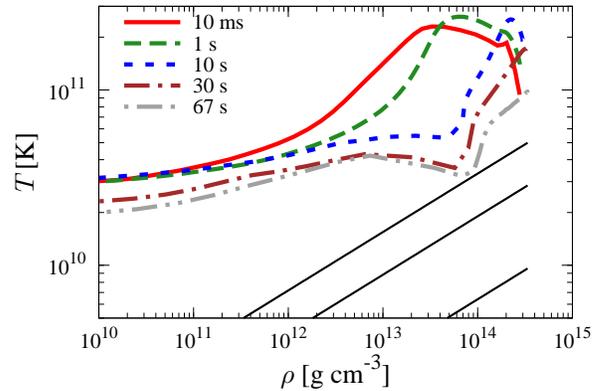}
\caption{The time evolution in the $\rho$-$T$ plane. The color and
  type of lines are the same as figure \ref{fig:hydro}. Three thin
  solid black lines indicate the critical lines for the crust
  formation. See text for details.  }
\label{fig:d-t}
\end{figure}

Figure \ref{fig:d-t} represents the time evolution in $\rho$-$T$
plane. The line colors and types are the same as figure
\ref{fig:hydro}.  In this plane, we show three black solid lines that
indicate the criteria for the crust formation.  The critical
temperature of lattice structure formation is given by \citet{shap83},
as
\begin{eqnarray}
T_{c}&\approx& \frac{Z^{2} e^{2}}{\Gamma k_{B}}\left(\frac{4\pi}{3}\frac{\rho Y_{e}x_a}{Zm_{u}}\right)^{{1/3}}\\
&\approx& 6.4\times 10^{9}~\mathrm{K}
\left(\frac{\Gamma}{175}\right)^{-1}
\left(\frac{\rho}{10^{14}~\mathrm{g~cm}^{-3}}\right)^{1/3}
\nonumber\\
&&\times
\left(\frac{Y_{e}}{0.1}\right)^{1/3}
\left(\frac{x_a}{0.3}\right)^{1/3}
\left(\frac{Z}{26}\right)^{5/3},
\label{eq:2}
\end{eqnarray}
where $Z$ is the typical proton number of the component of the
lattice, $e$ is the elementary charge, $\Gamma$ is a dimensionless
factor describing the ratio between the thermal and Coulomb energies
of lattice at the melting point, $k_{B}$ is the Boltzmann constant,
$x_a$ is the mass fraction of heavy nuclei, and $m_{u}$ is the atomic
mass unit, respectively. Critical lines are drawn using parameters of
$\Gamma=175$ (see, e.g., \cite{cham08}), $Y_e=0.1$, and $x_a=0.3$. As
for the proton number, we employ $Z$=26, 50, and 70 from bottom to
top.  Although the output for the typical proton number by the
equation of state is between $\sim$ 30 and 35, there is an objection
that the average proton number varies if we use the NSE
composition. In \citet{furu11}, they represented the mass fraction
distribution in the neutron number and proton number plane and implied
that even larger (higher proton number) nuclei can be formed in the
thermodynamic quantities considered here. Therefore, we here
parametrize the proton number and show the different critical lines
depending on the typical species of nuclei.  In addition, there are
several improved studies about $\Gamma$ that suggest the larger value
(e.g., \cite{horo07}), which leads to the lower critical temperature
corresponding to later crust formation, although the value is still
under debate.

The critical lines imply that the lattice structure is formed at the
point with the density of $\sim 10^{13-14}$ g cm$^{-3}$ and on the
postbounce time of $\sim 70$ s. Of course these values (especially the
formation time) strongly depend on employed parameters, but the
general trend would not change very much even if we include more
sophisticated parameters.

%\clearpage
%%%%%%%%%%%%%%%%%%%%%%%%%%%%%%%%%%%
%%%%%%%%%%%%%%%%%%%%%%%%%%%%%%%%%%%
\section{Summary and discussion}
\label{sec:summary}
%%%%%%%%%%%%%%%%%%%%%%%%%%%%%%%%%%%
%%%%%%%%%%%%%%%%%%%%%%%%%%%%%%%%%%%

In this letter, we performed a very long term simulation of the
supernova explosion for an 11.2 $M_{\odot}$ star. This is the first
simulation of an iron core starting from core collapse and finishing
in the PNS cooling phase. We focused on the PNS cooling phase by
continuing the neutrino-radiation-hydrodynamic simulation up to $\sim$
70 s from the onset of the core collapse. By comparing the thermal
energy and the Coulomb energy of the lattice, we finally found that
the temperature decreases to $\sim 3\times 10^{10}$ K with the density
$\rho\sim 10^{14}$ g cm$^{-3}$, which almost satisfies the critical
condition for the formation of the lattice structure. Even though
there are still several uncertainties for this criterion, this study
could give us informative inspection for the crust formation of a
NS. We found that the crust formation would start from the point with
$\rho\approx 10^{13-14}$ g cm$^{-3}$ and it would precedes from inside
to outside, because the Coulomb energy strongly depends on the mean
interstice between components so that the higher density exhibits the
earlier formation.

Next, we comment on our assumptions in this study. We performed 2D
simulation for the explosion phase and following 1D simulation for the
neutrino-driven wind phase and PNS cooling phase. It is well known
that the convection and SASI activity are different between 2D and 3D
in the explosion phase so that the 3D simulation should be performed
(see \cite{taki13},\cite{couc13}). In addition, in the PNS cooling phase we
observed the negative entropy gradient and negative lepton fraction
gradient as functions of the radius, which indicate that these regions
are convectively unstable so that in multi-D simulation the convective
motion could change the evolution quantitatively. However, it is still
too computationally expensive to perform 3D simulation with neutrino
radiative transfer up to 70 s postbounce and we think that our
findings in this work will not change in qualitative sense even when
the 3D simulations for such a long term is available in the
future. The possible direction in multi-D simulation for the crust
formation is the following; the convective motion could transfer the
heat from inside to outside more efficiently than the radiation so
that the temperature at the surface of PNS should be higher compared
to 1D run at the early time.  However, for the late time, the neutrino
luminosity could become smaller with convection than without
convection (see, e.g., \cite{robe12}).  This means that the surface
temperature decreases faster if we consider the convection, which
could lead to the faster formation of the crust.  In order to obtain
the final answer of the convective effects for the crust formation,
the long term evolution with multi-D hydrodynamic simulation is
strongly required, which is beyond the scope of this letter and will
be presented in forthcoming papers.

In addition to the dimensionality, we employed several simplification
in this study, i.e., the specific nuclear equation of state (EOS) by
\citet{shen98}, the simplified neutrino interactions based on
\citet{brue85}, the simplified neutrino transfer scheme by isotropic
diffusion source approximation (IDSA; \cite{lieb09}), and neglecting
the general relativistic (GR) effects. These treatments would lead to
quantitative difference of the crust formation time, so that more
sophisticated simulations are necessary to obtain more realistic
value.  In the following, we briefly discuss the possible direction by
improving these physics one by one.
% EOS
First, EOS affects the structure of PNS, especially its radius. Shen
EOS, which is used in this study, predicts $\sim$ 15 km for a cold NS
with 1.4 $M_{\odot}$, which is relatively larger than the suggested
value obtained by analysis of X-ray binaries (e.g.,
\cite{stei10}). The different EOS, which exhibits a smaller NS radius,
would result in the higher temperature of the crust formation and
later formation of the crust.
% neutrino interactions
Secondly, weak interactions used in this study are based on
\citet{brue85}, in which some interactions playing an important role
at PNS cooling phase are missing, e.g., nucleon-nucleon
bremsstrahlung. Therefore, our simulation could underestimate the
cooling rate and overestimate the temperature. Due to this
underestimation of the cooling, the shrinkage of the PNS is rather
slow in this study (several seconds; see figures \ref{fig:mass_shell}
and \ref{fig:hydro}) compared to previous studies ($\lesssim 1$ s,
see, e.g., \cite{fisc10}). In addition, some missing interactions might
cause the dip in $\rho$-$T$ plane shown in figure \ref{fig:d-t}. We
can expect that more improved simulations including these interactions
might lead to the earlier formation of the crust.
% IDSA
Thirdly, IDSA is known to produce considerable error around the
decoupling layer between the optically thick and thin regions
\citep{bern12}. This error comes from the simple description of IDSA,
in which we naively decompose the distribution function of neutrinos
into trapped part and free-streaming part \citep{lieb09}. This
treatment significantly simplify the transport equations for each
limit and can make the neutrino-radiation hydrodynamic simulations
computationally reasonable expense, so that we can perform such a
long-term simulation using the Eulerian-grid based code. However,
since this simplification exhibits error at the transition layer,
which determines the boundary condition of diffusion equation for
neutrinos and might change the thermal evolution even in the deep
core, more sophisticated transport scheme is necessary to obtain more
realistic time of the crust formation.
% GR
Finally, GR effects, which are neglected in this study, may change the
crust formation time because the stronger gravity exhibits a more
compact NS. More compact PNSs have higher entropy, which moves points
in upper-left direction slightly in $\rho$-$T$ plane so that the GR
effects are expected to delay the crust formation.
At last, it should be noted that even though the current study is
based on these simplifications expressed above, we believe that
qualitative features obtained here will not change dramatically even
if we include more sophisticated physics, i.e., the crust forms at
$\sim$10--100 s after the PNS formation near the center and the
crust-formation front propagates from inside to outside.

\bigskip

We thank the referee for providing constructive comments and help in
improving the contents of this letter, and M. Liebend\"orfer,
T. Muranushi, K. Sumiyoshi, M. Takano, and N. Yasutake for stimulating
discussion. Numerical computations in this study were in part carried
on XT4 and XC30 at CfCA in NAOJ and SR16000 at YITP in Kyoto
University. This study was supported in part by the Grants-in-Aid for
the Scientific Research from the Ministry of Education, Science and
Culture of Japan (No. 23840023 and 25103511) and HPCI Strategic
Program of Japanese MEXT.

\end{document}